\documentclass[conference, a4paper]{IEEEtran}
\IEEEoverridecommandlockouts
\usepackage{cite}
\usepackage{amsmath,amssymb,amsfonts}
\usepackage{algorithmic}
\usepackage{graphicx}
\usepackage{textcomp}
\usepackage{xcolor}
\usepackage{multirow}
\usepackage{balance}
\usepackage{color}
\usepackage{subfigure}
\usepackage{booktabs}
\usepackage{etoolbox}
\usepackage{siunitx}
\def\BibTeX{{\rm B\kern-.05em{\sc i\kern-.025em b}\kern-.08em
    T\kern-.1667em\lower.7ex\hbox{E}\kern-.125emX}}

\newcommand{\Fix}[1]{\textcolor{black}{#1}}
\begin{document}

\title{Adversarial Test on Learnable Image Encryption\\
}

\author{\IEEEauthorblockN{MaungMaung AprilPyone, Warit Sirichotedumrong and Hitoshi Kiya}
\IEEEauthorblockA{\textit{Department of Computer Science, Graduate School of System Design} \\
\textit{Tokyo Metropolitan University}\\
Asahigaoka, Hino-shi, Tokyo, 191-0065, Japan \\
\{april-pyone-maung-maung@ed.,warit-sirichotedumrong@ed.,kiya@\}tmu.ac.jp}
}

\maketitle

\begin{abstract}

  Data for deep learning should be protected for privacy preserving.
  Researchers have come up with the notion of learnable image encryption
  to satisfy the requirement. However,
  existing privacy preserving approaches have never considered the threat of
  adversarial attacks. In this paper, we ran an adversarial test
  on learnable image encryption in five different scenarios. The
  results show different behaviors of the network in the variable
  key scenarios and suggest learnable image encryption provides
  certain level of adversarial robustness.
\end{abstract}

\begin{IEEEkeywords}
learnable image encryption, adversarial robustness
\end{IEEEkeywords}

\section{Introduction}
Deep learning has brought major breakthroughs in computer vision as well as
other fields~\cite{Lecun2015}.
There is no doubt that it is due to better algorithms, bigger data and faster
computing resources.
With large amount of data, deep learning is often carried out in the cloud environments where privacy issues are generated.
To securely transmit images through an untrusted channel, researchers have proposed Encryption-then-Compression (EtC) systems such as~\cite{Chuman2019,sirichotedumrong_kiya_2019,kurihara2015encryption}.
Nevertheless, most of the traditional image encryption schemes are not compatible with deep learning.
Recently, a learnable image encryption method where a neural network can
learn encrypted images~\cite{Tanaka2018} and pixel-based image encryption designed for deep neural networks~\cite{WaritICIP2019} were proposed.

Although privacy issues have been addressed, the security of deep learning has never been considered in the context of learnable image encryption.
Adversarially robust models are extremely desired for privacy-preserving schemes because the protected data are usually sensitive.
Unexpected misclifications due to adversarial attacks will lead to severe damage to applications such as medical analysis, surveillance, etc.
Therefore, security is quintessential in privacy-preserving deep neural networks.
In the literature, it has been shown that state-of-the-art neural networks are vulnerable towards adversarial examples~\cite{Szegedy2013, Biggio2013}.
Adversarial examples can be generated by optimization techniques to maximize the loss such as~\cite{Goodfellow2014, Kurakin2016, Carlini2017}.
Since then, deep learning has got significant amount of attention towards adversarial robustness~\cite{Madry2017}.

Our hypothesis is that encryption for privacy protection can give us somewhat
adversarial robustness. In this paper, we run a test on a recent learnable
encryption method~\cite{Tanaka2018} where input images are encrypted and sent
to the network with an adaptation layer. From our experiments, we confirm that
learnable image encryption has somewhat resistance to adversarial examples.
Our contribution in this paper is that we conducted a test to raise
a fundamental need of adversarial robustness in the privacy-preserving network.
We are also the first to consider adversarial perspective in learnable image
encryption.

\begin{figure}[t]
  \centerline{\includegraphics[width=2in]{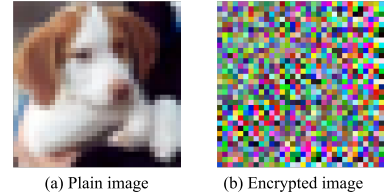}}
\caption{An example of encrypted image.\label{fig:encrypted}}
\end{figure}

\section{Preliminaries}
\subsection{Learnable Image Encryption}
Fig.~\ref{fig:encrypted} illustrates an example of encrypted image by Tanaka's encryption.
There are 2 parts in Tanaka's work: image encryption and adaptation layer~\cite{Tanaka2018}.

\begin{figure}[t]
  \centerline{\includegraphics[width=\linewidth]{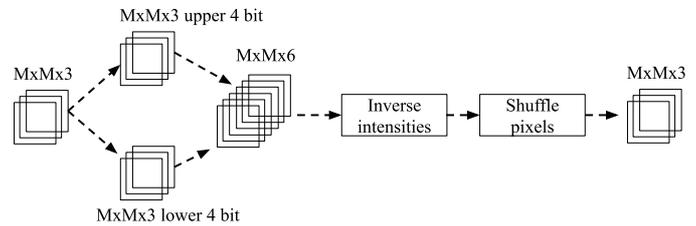}}
  \caption{Encyption process of a block by Tanaka's method~\cite{Tanaka2018}.\label{fig:tanaka}}
\end{figure}

The 8-bit pixel values in MxM blocks are separated into upper and lower 4-bit to form the 6-channel blocks.
The intensities of pixel values are randomly reversed and shuffled.
The 6-channel blocks are reformed to 3-channel blocks. For simplicity, the encryption process of one block is depicted in Fig.~\ref{fig:tanaka}.

\begin{figure}[t]
  \centerline{\includegraphics[width=\linewidth,height=\textheight,keepaspectratio]{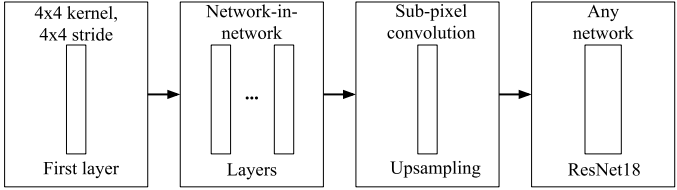}}
  \caption{Diagram of adaptation network by Tanaka's method~\cite{Tanaka2018}.\label{fig:tanaka-adapt}}
\end{figure}

The adaptation comprises of the first convolution layer (MxM kernel and MxM
stride), several network-in-network style layers and sub-pixel
convolution (pixel shuffle). After the adaptation network, any network can be followed.
Fig.~\ref{fig:tanaka-adapt} describes the diagram of Tanaka's adaptation network.


\subsection{Adversarial Examples}
\begin{figure}[t]
  \centerline{\includegraphics[width=3in]{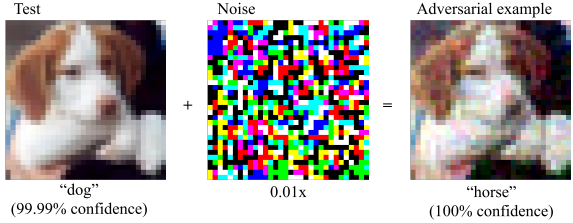}}
\caption{A sample of adversarial example generated by PGD.\label{fig:adversarial-example}}
\end{figure}

Adversarial examples are perceptually same images as the original ones that are
carefully designed to make the neural network misclassify with high confidence.
Fig.~\ref{fig:adversarial-example} shows an example of adversarial example
where the network classifies ``dog'' as ``horse'' with \SI{100}{\percent} confidence.

There are many different ways of crafting adversarial examples.
The popular and computationally efficient one is known as
Fast Gradient Sign Method (FGSM)~\cite{Goodfellow2014}. In this work, we
consider a stronger adversary (i.e., multi-step FGSM) known as projected
gradient descent (PGD)~\cite{Kurakin2016}:
\begin{equation}
  x^{t+1} = clip_{\epsilon}(x^{t} + \alpha \: sign(\nabla_{x}L(\theta,x,y))), \label{eq:pgd}
\end{equation}
where $\epsilon$ is allowable perturbation size, $x^t$ is adversarial
example on $t$\textsuperscript{th} PGD iteration, $\alpha$ is step size and $\nabla_{x}L(\theta,x,y)$ represents the gradients of the loss function.
Instead of multiplying the sign of the gradients with $\epsilon$ directly, PGD only adds some step size $\alpha$ to the image in each iteration.
PGD also projects the perturbation back to the max-norm in each step (i.e., $clip_{\epsilon}(.)$).

\subsection{Adversarial Training}
Adversarial training is to train a network to be robust against adversarial
examples. There are many types of adversarial defense. Madry et al.\ show that
training against PGD leads to robustness against other first-order
adversaries~\cite{Madry2017}. Therefore, we focus on PGD training in this work for
experimental purposes.

\section{Experiments}
\begin{figure}[t]
  \centerline{\includegraphics[width=\linewidth,height=\textheight,keepaspectratio]{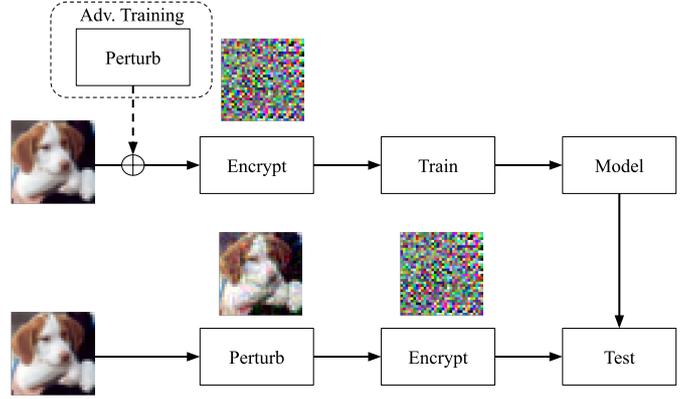}}
  \caption{The flow diagram of the experiment (``Adv. Training'' refers to ``Adversarial Training'' ).\label{fig:gcce-system}}
\end{figure}

The flow diagram of the experiment is presented in Fig.~\ref{fig:gcce-system}.
First, we encrypted the data and used the encrypted images to train the network by reusing the encryption code released by the author in GitHub~\cite{tanakagithub}.
For the testing, We generated adversarial noise by PGD, added the noise to the test images and then, encrypted them.

The experiment setup is described as follows. We used CIFAR10~\cite{Krizhevsky2009} dataset with batch size of 128.
Then, we implemented Tanaka's adaptation layer on deep residual network (ResNet18)~\cite{He2016} on PyTorch platform.
The network was trained for 100 epochs with stochastic gradient descent optimization on learning rate of 0.1.
We reduced the learning rate to 0.01 after running for 40 epochs. All the images in the dataset for all cases were in the
range of $[0, 1]$ with live augumentation (random cropping and random
horizontal flip). However, there was no prior normalization.
For the adversarial testing and training, we used PGD~\cite{Kurakin2016} with $\epsilon=0.1$,
$\alpha=0.01$ for 20 iterations.

\Fix{Adversarial noise was generated iteratively as described in~\eqref{eq:pgd}.
We used a slightly higher noise level (i.e., $\epsilon = 0.1$) to stress the experiment.
The PGD process with the same settings was employed in both adversarial testing and training.}

We trained the network in 5 different scenarios and tested against adversarial examples generated by PGD\@.
\begin{enumerate}
  \item Plain: ResNet18 trained on clean images
  \item Encrypted: ResNet18 with Tanaka's adaptation layer trained on encrypted images (one key for both training and testing)
  \item Encrypted-Adv: Adversarial training for encrypted images using the same key
  \item Encrypted-DK:\@ Same model as \textbf{Encrypted} except each batch of the dataset was encrypted using a different random key
  \item Encrypted-Adv-DK:\@ Adversarial training for encrypted images using different random keys
\end{enumerate}

\section{Results}
Table~\ref{tab:results} summarizes the results of the experiments.
The error is the accuracy of misclassification (the lower the better).
We recorded the errors during training, testing and adversarial testing.
The results suggest the followings:

\begin{enumerate}
  \item The standard model trained with plain images are completely vulnerable towards
    adversarial examples (i.e., \SI{100}{\percent} misclassification).
  \item Image encryption provides certain degree of robustness against adversarial examples.
    The error rate of 0.269 is extremely good for adversarial robustness.
    Even the state-of-the-art adversarially trained model had the approximate
    error rate of 0.530 for PGD attack with 20 steps~\cite{cifar10challenge}.
    This confirms our hypothesis that the encryption can give certain level of
    adversarial robustness.
  \item Traditional adversarial training is not suitable for learnable image encryption.
    While doing adversarial training with the same key, the network became biased towards adversarial examples.
    The network could not generalize the clean examples. Therefore, the network performed poor on the test dataset.
  \item The neural network requires uniform encryption for both train and test data.
    Different keys transform the data into different distributions.
    Therefore, when using different keys for encryption, the network did not learn.
  \item Since the network assumes encryption with one key, adversarial training with different keys is not relevant.
    We carried out the test for experimental purposes only.
\end{enumerate}

\section{Conclusions}
Learnable image encryption provides good adversarial robustness.
The same key is necessary to uniformly encrypt the images.
In addition, we can also conclude that the standard adversarial training is not suitable for encrypted images.
We aim to achieve secure and private models with reasonable accuracy as our
future work.

\robustify\bfseries
\sisetup{table-parse-only,detect-weight=true,detect-inline-weight=text,round-mode=places,round-precision=3}

\begin{table}[tbp]
  \caption{Adversarial Test Results\label{tab:results}}
  \centering
  \begin{tabular}{lSSS}
  \toprule
  \multirow{2}{*}{Model}&\multicolumn{3}{c}{Error} \\
  \cmidrule{2-4} 
  & {Train} & {Test} & {Adversarial} \\
  \midrule
  Plain & 0.006 & 0.137 & 1.000\\
  Encrypted & \bfseries \num{0.011} & \bfseries \num{0.166} & \bfseries \num{0.269}\\
  Encrypted-Adv$^{\dagger}$ & 0.09340 & 0.829500 & 0.039600\\
  Encrypted-DK$^{\ddagger}$ & 0.431 & 0.506 & 0.792\\
  Encrypted-Adv$^{\dagger}$-DK$^{\ddagger}$ & 0.657 & 0.617 & 0.690\\
  \bottomrule

  \multicolumn{4}{l}{$^{\dagger}$Adversarial, $^{\ddagger}$Diffferent keys.}
  \end{tabular}
\end{table}


\end{document}